\documentclass[10pt,journal,twocolumn]{IEEEtran}
\usepackage{cite}
\usepackage{amsmath,amssymb,amsfonts}
\usepackage{graphicx}
\usepackage{textcomp}
\usepackage{amsthm}
\usepackage{optidef}

\newtheorem{theorem}{Theorem}[section]
\newtheorem{lemma}[theorem]{Lemma}
\newtheorem{remark}{Remark}

\usepackage{multicol, blindtext}
\usepackage{lipsum}
\usepackage{afterpage,lipsum,capt-of,graphicx}
\usepackage{caption}
\usepackage{cuted}
\usepackage{algorithmicx}
\usepackage{algpseudocode}
\usepackage{algorithm}
\algblockdefx{MRepeat}{EndRepeat}{\textbf{repeat}}{}
\algnotext{EndRepeat}
\enlargethispage{-2cm}
\usepackage{mathtools}
\usepackage{pbox}
\usepackage{mathtools}
\usepackage{subcaption}
\usepackage{refcount}
\usepackage{flushend}

\usepackage{relsize}
\begin{document}

\title{\Large \bf Event-Based Distributed Linear Quadratic Gaussian for Multi-Robot Coordination with
Localization Uncertainty}

\author{Tohid Kargar Tasooji, Sakineh Khodadadi

\thanks{ Tohid Kargar Tasooji is with the Department of Aerospace Engineering,
Toronto Metropolitan University, Toronto, ON M5B 2K3, Canada (e-mail:
tohid.kargartasooji@torontomu.ca).  Sakineh Khodadadi is with the Department of Electrical and Computer
Engineering, University of Alberta, Edmonton,AB,  AB T6G 1H9, Canada (email: sakineh@ualberta.ca).}
}

\maketitle

\begin{abstract} \small\baselineskip=9pt
This paper addresses the problem of event-based distributed Linear Quadratic Gaussian (LQG) control for multi-robot coordination under localization uncertainty. An event-triggered LQG rendezvous control strategy is proposed to ensure coordinated motion while reducing communication overhead. The design framework decouples the LQG controller from the event-triggering mechanism, although the scheduler parameters critically influence rendezvous performance. We establish stochastic stability for the closed-loop multi-robot system and demonstrate that a carefully tuned event-triggering scheduler can effectively balance rendezvous accuracy with communication efficiency by limiting the upper bound of the rendezvous error while minimizing the average transmission rate. Experimental results using a group of Robotarium mobile robots validate the proposed approach, confirming its efficacy in achieving robust coordination under uncertainty. 
\end{abstract}


\section{Introduction} \IEEEPARstart{I}{N} recent years, cooperative control for multi-robot systems has attracted significant attention due to its extensive applications in traffic management, search and rescue operations, formation control, and many other fields \cite{1,2,3,4, 25, 26, 27, 28, 29, 30, 36, 37, 38, 39}. A critical task in these systems is achieving rendezvous, where all robots converge to a common state through local interactions. Traditionally, continuous state exchange has been employed to achieve rendezvous, but this approach often leads to excessive consumption of communication resources, such as bandwidth and energy.

To address these issues, event-triggered control has emerged as a promising alternative, enabling robots to update and share their state information only when necessary, as dictated by a pre-specified event condition. Several studies have explored event-triggered rendezvous control. For instance, in \cite{11} an integral-type event-triggering condition was introduced for asynchronous distributed systems, while \cite{12} developed an event-triggered strategy for two-wheeled mobile robots under time-varying communication delays. Additional approaches have been proposed for single-integrator networks \cite{13}, multi-robot systems with input saturation \cite{14}, heterogeneous dynamics \cite{15}, adaptive leader-following scenarios \cite{16}, and nonlinear second-order systems \cite{17,18}.

Despite these advances, many existing works ignore the impact of sensor noise and localization uncertainty, which are inherent in practical robotic applications. In realistic settings, noisy sensor data necessitates the use of robust estimation techniques such as the Kalman filter. Moreover, while event-triggered schemes reduce communication load, a critical challenge remains in jointly optimizing the control performance and the communication rate.

Motivated by these challenges, this paper presents an event-based distributed Linear Quadratic Gaussian (LQG) rendezvous control strategy for multi-robot systems operating under localization uncertainty. The key contributions of this work are as follows:

\begin{itemize} \item \textbf{Distributed LQG Rendezvous Control:} We propose an optimal event-triggered LQG control framework that simultaneously addresses the trade-off between control performance and communication efficiency. Unlike prior studies that neglect measurement noise, our framework incorporates noisy sensor data through Kalman filtering. A novel mixed-type event-triggering condition is introduced, allowing each robot to decide whether to transmit its state estimate to its neighbors at each time step. This approach not only reduces the communication burden but also avoids Zeno behavior in the presence of noise. \item \textbf{Stochastic Stability Analysis:} We provide a rigorous stochastic stability analysis of the proposed rendezvous control scheme. Our analysis shows that by appropriately tuning the event-triggering scheduler, the system can achieve satisfactory rendezvous performance while significantly reducing the average transmission rate. \item \textbf{Experimental Validation on a Robotrium Platform:} To demonstrate the practical effectiveness of the proposed control strategy, we implement it on a multi-robot platform using Robotrium mobile robots. Unlike previous studies that rely primarily on simulations, our experimental setup addresses real-world challenges, such as communication delays, limited battery capacity, and mechanical constraints. We compare various triggering conditions and show that the proposed mixed-type strategy achieves an optimal balance between rendezvous performance and communication efficiency. \end{itemize}

The remainder of this paper is organized as follows. Section II formulates the problem and reviews relevant preliminaries. Section III details the design of the event-triggered LQG rendezvous control law and provides the stability analysis. Section IV presents the experimental setup and results from tests on the Robotrium platform. Finally, Section V concludes the paper and outlines directions for future research.

\section{Problem formulation and Preliminaries}
\subsection{Notation}
$\mathbb{R}^n$ represents the $n$-dimensional vectors in Euclidean space, $\mathbb{R}^{n \times m}$ denotes the matrices with $n$ rows and $m$ columns, $\left\| x \right\|$ stands for the Euclidean norm of the vector $x\in \mathbb{R}^n$. $X^T \in \mathbb{R}^{m \times n}$ denotes its transpose of the matrix $X \in \mathbb{R}^{n \times m}$. $I_{n}$ is the identity matrix of $n$ dimensions. $\mathrm{Tr}(X)$ is the trace of matrix $X$. For matrices $A \in \mathbb{R}^{m \times n}$ and  $B \in \mathbb{R}^{p \times q}$, their Kronecker product is a matrix in $ \mathbb{R}^{pm \times qn}$  represented as $A \otimes B$. 
 The notation $X > 0$ (respectively $X \ge 0$) means that the matrix $X \in \mathbb{R}^{n \times n}$  is positive definite (respectively positive semi-definite), {\it i.e.}, all its eigenvalues
are positive or nonnegative. 

We consider a MAS with $N$ agents which communicate with each other according to the interaction topology denoted by graph $\mathcal{G}=(\mathcal{V},\mathcal{E}, \mathcal{A})$. Here $\mathcal{V}=\{v_{1},...,v_{N} \}$ is the set of nodes in the graph. For a node $v_{i}$, the agent index $i \in \{1,2,...,N \}$ is the unique identifier of the agent $i$. $\mathcal{E}$ represents the edge set, {\it i.e.} $v_{ij}=(v_{i},v_{j}) \in \mathcal{E}$ if agent $i$ received information from agent $j$. $\mathcal{A}=\lfloor a_{ij}\rfloor \in \mathbb{R}^{N \times N}$ is the adjacency matrix where $a_{ij}=1$ if $v_{ij} \in \mathcal{E}$ otherwise $a_{ij}=0$. Also, $a_{ii}=0$ $\forall	i \in \{1,2,...,N \}$. The graph Laplacian $\mathcal{L}=[l_{ij}] \in \mathbb{R}^{n_{i} \times n_{i}} $ is defined as $l_{ij}=\sum_{i\neq j} a_{ij}$ and $l_{ij}=-a_{ij}$  where $i\neq j$.

\subsection{Dynamics of 2D Mobile Robots}

Consider a team of \( N \) mobile robots, where the state of each robot \( i \in \{1,2,\ldots,N\} \) at time step \( k \) is defined by its position in the plane:
\[
x_k^i = \begin{bmatrix} x_k^i \\ y_k^i \end{bmatrix} \in \mathbb{R}^{2}.
\]
The evolution of the position is modeled by a discrete-time stochastic linear system:
\begin{equation}   \label{eq:robot_dynamics_position}
x_{k+1}^i = A x_k^i + B u_k^i + w_k^i,
\end{equation}
where \( u_k^i \in \mathbb{R}^{2} \) represents the control input (typically the velocity command or desired displacement in the \( x \) and \( y \) directions), \( A \in \mathbb{R}^{2 \times 2} \) and \( B \in \mathbb{R}^{2 \times 2} \) are the system and input matrices respectively, and \( w_k^i \) denotes the process noise. The noise \( w_k^i \) is assumed to be a zero-mean Gaussian white noise with covariance \( W^i \in \mathbb{S}^{2} \).

The initial state \( x_0^i \) is modeled as a Gaussian random variable with mean \( \bar{x}_0^i \) and covariance \( X_0 \in \mathbb{S}^{2} \).

Due to sensor limitations, the state of each robot is not directly observable and is instead estimated from noisy measurements given by
\begin{equation}   \label{eq:robot_measurement_position}
y_k^i = C x_k^i + v_k^i,
\end{equation}
where \( y_k^i \in \mathbb{R}^{2} \) is the measurement vector, \( C \in \mathbb{R}^{2 \times 2} \) is the output matrix (often chosen as the identity matrix, i.e., \( C = I_2 \), when direct position measurements are available), and \( v_k^i \) is a zero-mean Gaussian white noise with covariance \( V^i \in \mathbb{S}^{2} \).

This simplified model, focusing solely on the planar position, serves as the foundation for the development of the event-triggered LQG rendezvous control strategy, ensuring effective coordination among 2D mobile robots under localization uncertainty.
\subsection{Kalman Filtering for 2D Mobile Robots}

In practical scenarios, the state of each 2D mobile robot, representing its position in the plane, is measured with noise. To obtain accurate position estimates from noisy sensor data, we employ a Kalman filter, which recursively computes the minimum mean squared error (MMSE) estimates. These state estimates are subsequently communicated to neighboring robots over the network. Note that transmitting the processed state estimates generally yields better performance compared to sharing raw sensor measurements \cite{23}.

Let 
\[
\mathcal{D}_k^i \triangleq \{u_0^i, u_1^i, \ldots, u_k^i; \, y_0^i, y_1^i, \ldots, y_k^i\}
\]
denote the history of control inputs and measurements for robot \( i \) up to time \( k \). The Kalman filter operates in two key steps:

\textbf{1) Prediction (Propagation) Step:}  
Based on the previous estimate \(\hat{x}_{k-1|k-1}^i\), the state is predicted as
\begin{equation}   \label{eq:kalman_predict_state}
\hat{x}_{k|k-1}^i = A \hat{x}_{k-1|k-1}^i + B u_{k-1}^i,
\end{equation}
where \(A\) and \(B\) are the system and input matrices from the 2D mobile robot dynamics. The prediction error is defined by
\begin{equation}   \label{eq:kalman_predict_error}
\tilde{x}_{k|k-1}^i \triangleq x_k^i - \hat{x}_{k|k-1}^i,
\end{equation}
with its associated covariance given by
\begin{equation}   \label{eq:kalman_predict_cov}
P_{k|k-1}^i = A P_{k-1|k-1}^i A^T + W^i,
\end{equation}
where \(W^i\) represents the process noise covariance.

\textbf{2) Update (Correction) Step:}  
Upon receiving the new measurement \(y_k^i\), the state estimate is updated as follows:
\begin{equation}   \label{eq:kalman_update_state}
\hat{x}_{k|k}^i = \hat{x}_{k|k-1}^i + K_k^i \left(y_k^i - C \hat{x}_{k|k-1}^i\right),
\end{equation}
where \(C\) is the measurement matrix and \(K_k^i\) denotes the Kalman gain. The updated estimation error becomes
\begin{equation}   \label{eq:kalman_update_error}
\tilde{x}_{k|k}^i \triangleq x_k^i - \hat{x}_{k|k}^i,
\end{equation}
with the corresponding covariance updated according to
\begin{equation}   \label{eq:kalman_update_cov}
P_{k|k}^i = \left(I - K_k^i C\right) P_{k|k-1}^i.
\end{equation}

The Kalman gain \(K_k^i\) is computed as
\begin{equation}   \label{eq:kalman_gain}
K_k^i = P_{k|k-1}^i C^T \left(C P_{k|k-1}^i C^T + V^i\right)^{-1},
\end{equation}
where \(V^i\) is the covariance matrix of the measurement noise.

This recursive filtering process enables each robot to continuously refine its position estimate, thereby enhancing the performance of the subsequent event-triggered rendezvous control strategy.

\subsection{Event-Triggered Communication Mechanism}

To reduce communication overhead and conserve energy in multi-robot networks, an event-triggered communication mechanism is employed. In this framework, each 2D mobile robot updates its local state estimate \(\hat{x}_{k|k}^i\) synchronously at every sampling instant \(k\), and a local scheduler determines whether this estimate should be transmitted to neighboring robots. A transmission occurs only when the update is deemed significant, thereby conserving network resources.

We consider two distinct event-triggered communication mechanisms (ETCMs):

\textbf{Mechanism 1 (Mixed-Type Triggering):}  
A transmission is initiated at time \(k\) if
\begin{equation}   \label{eq:etcm1}
\| e_k^i \|^2 \geq \alpha_k^i \| \hat{x}_{k|k}^i \|^2 + \beta_k^i,
\end{equation}
where the error is defined by
\[
e_k^i = \hat{x}_{k|k}^i - \hat{x}_{\tau_k|\tau_k}^i,
\]
with \(\hat{x}_{\tau_k|\tau_k}^i\) being the most recent state estimate that was broadcast. The design parameters \(\alpha_k^i\) and \(\beta_k^i\) tune the sensitivity of the trigger. In particular, setting \(\alpha_k^i = 0\) recovers a send-on-delta rule, while \(\beta_k^i = 0\) yields a relative triggering scheme. The timestamp \(\tau_k\) for the last transmission is updated as follows:
\begin{equation}   \label{eq:tau_update}
\tau_k = 
\begin{cases}
k, & \text{if } \hat{x}_{k|k}^i \text{ is transmitted}, \\
\tau_{k-1}, & \text{otherwise}.
\end{cases}
\end{equation}

\textbf{Mechanism 2 (Integral-Based Triggering):}  
An alternative approach leverages an integrated error measure over time. Here, a transmission is triggered at time \(k\) if
\begin{equation}   \label{eq:etcm2}
\sum_{j=\tau_k}^{k} \| e_j^i \| \, \| \hat{x}_{j|j}^i \| \geq \gamma_k^i \sum_{j=\tau_k}^{k} \| \hat{x}_{j|j}^i \|^2,
\end{equation}
where \(\gamma_k^i\) is a predetermined threshold parameter and \(\tau_k\) is as defined in \eqref{eq:tau_update}. This condition accumulates the magnitude of the estimation errors over time and triggers a transmission once the integrated error exceeds a fraction of the cumulative state magnitude.

For performance evaluation, the average transmission rate for robot \(i\) is defined as
\begin{equation}   \label{eq:transmission_rate}
\Gamma^i = \lim_{T\to\infty} \sup \frac{1}{T} \mathbb{E} \left[ \sum_{k=0}^{T-1} \sigma_k^i \right],
\end{equation}
where \(\sigma_k^i\) equals 1 if a transmission occurs at time \(k\) and 0 otherwise. A value of \(\Gamma^i = 1\) corresponds to continuous (periodic) communication, while lower values indicate significant communication savings.

The proposed event-triggered communication mechanism thus provides an effective means to balance the trade-off between maintaining up-to-date state information and minimizing communication load in multi-robot coordination tasks.

\subsection{Event-Triggered Distributed Consensus Protocol and Performance Criterion}

In our framework, each 2D mobile robot implements a distributed consensus protocol that relies on the most recent state estimates received from its neighbors via an event-triggered communication mechanism. Specifically, the control input for robot \(i\) is defined as
\begin{equation}   \label{eq:consensus_protocol}
u_k^i = -L_k \sum_{j \in \mathcal{N}_i} \Bigl( \hat{x}_{\tau_k| \tau_k}^i - \hat{x}_{\tau_k| \tau_k}^j \Bigr),
\end{equation}
where \(\mathcal{N}_i\) denotes the set of neighboring robots for agent \(i\), \(\hat{x}_{\tau_k| \tau_k}^j\) represents the latest transmitted state estimate from robot \(j\), and \(L_k\) is the consensus gain matrix to be designed.

To quantify consensus performance, we introduce the local estimation error relative to the desired rendezvous point \(x_0\):
\begin{equation}   \label{eq:consensus_error}
\hat{\varepsilon}_k^i = \hat{x}_{k|k}^i - x_0.
\end{equation}
Our objective is to design a consensus protocol that minimizes both the deviation from the common rendezvous point and the communication cost incurred by event-triggered transmissions.

Toward this end, we consider a finite-horizon quadratic performance index for each robot \(i\) given by
\begin{equation}   \label{eq:performance_index}
J^i = \mathbb{E}\left[ \hat{\varepsilon}_M^{iT} Q_M^i \hat{\varepsilon}_M^i + \sum_{k=0}^{M-1} \left( \hat{\varepsilon}_k^{iT} Q^i \hat{\varepsilon}_k^i + u_k^{iT} R^i u_k^i \right) \right],
\end{equation}
where \(Q^i = Q^{iT} \succeq 0\) and \(Q_M^i = Q_M^{iT} \succeq 0\) are the state error weighting matrices, and \(R^i = R^{iT} \succ 0\) is the control input weighting matrix. In this formulation, minimizing \(J^i\) ensures convergence to the rendezvous point with desirable performance, while the overall system design also incorporates the objective of reducing the average transmission rate (as defined in \eqref{eq:transmission_rate}) through the event-triggered communication strategy.

The joint design of the consensus gain \(L_k\) and the event-triggering parameters enables a balanced trade-off between control performance and communication efficiency, thereby ensuring that the network of 2D mobile robots achieves robust and efficient rendezvous in the presence of measurement noise and limited communication resources.

\section{Main Results}

In this section, we present the main results on the design of an event-triggered distributed consensus protocol for a network of 2D mobile robots subject to stochastic disturbances. By combining the Kalman filter update, the consensus protocol, and the event-triggered communication mechanism, the global dynamics of the network can be written in compact form as

\begin{equation} \label{eq:global_dynamics}
\begin{aligned}
\hat{\bar{x}}_{k+1|k+1} &= (I_N \otimes A) \hat{\bar{x}}_{k|k} + (\mathcal{L} \otimes B) \bar{u}_k - (\mathcal{L} \otimes B L_k) \bar{e}_k \\
&\quad + \bar{K}_{k+1} \Bigl[ \bar{y}_{k+1} - (I_N \otimes C) \hat{\bar{x}}_{k+1|k} \Bigr],
\end{aligned}
\end{equation}
where the stacked vectors are defined as
\begin{equation*}
\hat{\bar{x}}_{k+1|k+1} =
\begin{bmatrix}
\hat{x}_{k+1|k+1}^{1} \\ \vdots \\ \hat{x}_{k+1|k+1}^{N}
\end{bmatrix}, \quad
\bar{u}_k =
\begin{bmatrix}
u_k^{1} \\ \vdots \\ u_k^{N}
\end{bmatrix}, \quad
\bar{K}_{k+1} =
\begin{bmatrix}
K_{k+1}^{1} \\ \vdots \\ K_{k+1}^{N}
\end{bmatrix},
\end{equation*}
and the measurement, prediction error, and triggering error vectors are similarly defined. Equation \eqref{eq:global_dynamics} is further recast as
\begin{equation} \label{eq:global_dynamics_recast}
\begin{aligned}
\hat{\bar{x}}_{k+1|k+1} &= (I_N \otimes A) \hat{\bar{x}}_{k|k} + (\mathcal{L} \otimes B) \bar{u}_k - (\mathcal{L} \otimes B L_k) \bar{e}_k \\
&\quad + \vartheta_k \eta_k,
\end{aligned}
\end{equation}
where
\begin{equation} \label{eq:theta_eta}
\vartheta_k = \Bigl[ \bar{K}_{k+1}(I_N \otimes C)(I_N \otimes A) \quad \bar{K}_{k+1}(I_N \otimes C) \quad \bar{K}_{k+1} \Bigr],
\end{equation}
and
\begin{equation} \label{eq:eta_def}
\eta_k = 
\begin{bmatrix}
\tilde{\bar{x}}_{k|k} \\ \bar{w}_k \\ \bar{v}_{k+1}
\end{bmatrix}.
\end{equation}

We next state a lemma that will be useful in our analysis.

\begin{lemma}[\cite{20}] \label{lemma:expectation}
Let \( x \) be a normally distributed vector with mean \( m \) and covariance \( R \). Then, for any stochastic matrix \( S \),
\begin{equation} \label{eq:lemma_expectation}
\mathbb{E}\{ x^T S x \} = m^T \, \mathbb{E}\{S\} \, m + \mathrm{tr} \bigl( \mathbb{E}\{S\} \, R \bigr).
\end{equation}
In the special case when \( S \) is constant, this simplifies to
\begin{equation} \label{eq:lemma_expectation_constant}
\mathbb{E}\{ x^T S x \} = m^T S m + \mathrm{tr}(S R).
\end{equation}
\end{lemma}

Based on the developments above, we now summarize the following main theorem.

\begin{theorem} \label{thm:main}
Consider the stochastic discrete-time system \eqref{eq:global_dynamics} with the event-triggered communication rules specified in Section II. Then, there exists a distributed consensus protocol of the form
\begin{equation} \label{eq:consensus_protocol_final}
\bar{u}_k = -L_k \hat{\bar{\varepsilon}}_{k|k},
\end{equation}
which minimizes the finite-horizon quadratic cost
\begin{equation} \label{eq:cost_function_global}
J = \mathbb{E}\Biggl\{ \hat{\bar{\varepsilon}}_M^T \bar{Q}_M \hat{\bar{\varepsilon}}_M + \sum_{k=0}^{M-1} \Bigl( \hat{\bar{\varepsilon}}_k^T \bar{Q} \hat{\bar{\varepsilon}}_k + \bar{u}_k^T \bar{R} \bar{u}_k \Bigr) \Biggr\},
\end{equation}
with \(\hat{\bar{\varepsilon}}_k = \hat{\bar{x}}_{k|k} - \mathbf{1}_N \otimes x_0\) denoting the consensus error (relative to the rendezvous point \( x_0 \)). The optimal gain matrix \( L_k \) is given by
\begin{equation} \label{eq:optimal_gain}
\begin{aligned}
L_k = \Bigl[ \bar{R} + \mathbb{E} \{ (\mathcal{L} \otimes B)^T \Sigma_{k+1} (\mathcal{L} \otimes B) \} \Bigr]^{-1} \\ \times \mathbb{E} \{ (\mathcal{L} \otimes B)^T \Sigma_{k+1} (I_N \otimes A) \},
\end{aligned}
\end{equation}
and the matrix sequence \( \{\Sigma_k\} \) satisfies the recursion
\begin{equation} \label{eq:riccati_recursion}
\begin{split}
\Sigma_k &= \mathbb{E} \Bigl\{ \Bigl[ (I_N \otimes A) - (\mathcal{L} \otimes B)L_k \Bigr]^T \\& \times \Sigma_{k+1}  \Bigl[ (I_N \otimes A)  - (\mathcal{L} \otimes B)L_k \Bigr] \Bigr\} 
 + L_k^T \bar{R} L_k + \bar{Q},
\end{split}
\end{equation}
with terminal condition
\begin{equation} \label{eq:riccati_terminal}
\Sigma_M = \bar{Q}_M.
\end{equation}
The minimum cost achieved by the optimal protocol is expressed as
\begin{equation} \label{eq:optimal_cost}
\begin{aligned}
J_k &= \hat{\bar{\varepsilon}}_M^T \Sigma_M \hat{\bar{\varepsilon}}_M + \mathrm{tr}(\Sigma_M \bar{X}_M) \\
&\quad + \sum_{k=0}^{M-1} \Biggl\{ \hat{\bar{\varepsilon}}_k^T \Bigl[ \mathbb{E} \Bigl\{ \Bigl( (I_N \otimes A) - (\mathcal{L} \otimes B)L_k \Bigr)^T \Sigma_{k+1} \\
&\quad \quad \times \Bigl( (I_N \otimes A) - (\mathcal{L} \otimes B)L_k \Bigr) \Bigr\} + L_k^T \bar{R} L_k + \bar{Q} \Bigr] \hat{\bar{\varepsilon}}_k \\
&\quad + \mathrm{tr} \Bigl\{ (I_N \otimes A)^T (I_N \otimes C)^T \bar{K}_{k+1}^T \Sigma_{k+1} \bar{K}_{k+1} (I_N \otimes C) \\
&\quad \quad \times (I_N \otimes A) \bar{P}_{k|k} \Bigr\} + \mathrm{tr}\bigl( \bar{K}_{k+1}^T \Sigma_{k+1} \bar{K}_{k+1} \bar{V} \bigr) \\
&\quad + \mathrm{tr}\Bigl( (I_N \otimes C)^T \bar{K}_{k+1}^T \Sigma_{k+1} \bar{K}_{k+1} (I_N \otimes C) \bar{W} \Bigr) \\
&\quad + \mathrm{tr}\Bigl\{ (\mathcal{L} \otimes B L_k)^T \Sigma_{k+1} (\mathcal{L} \otimes B L_k) \mathbb{E} \{ \bar{e}_k^T \bar{e}_k \} \Bigr\} \Biggr\}.
\end{aligned}
\end{equation}
\end{theorem}

\begin{proof}
The proof is based on a dynamic programming argument. Let 
\[
D_k \triangleq \Bigl\{ \bar{u}_0, \bar{u}_1, \ldots, \bar{u}_k; \, \bar{y}_0, \bar{y}_1, \ldots, \bar{y}_k \Bigr\}
\]
denote the information available up to time \(k\). Exploiting the fact that the conditional expectation of the error given \( D_k \) is fully characterized by the local state estimate \( \hat{\bar{\varepsilon}}_{k|k} \) (see \cite{21}), the Bellman equation is formulated as
\begin{equation} \label{eq:bellman}
J_k = \min_{\bar{u}_k} \mathbb{E} \Bigl\{ \bar{\varepsilon}_k^T \bar{Q} \bar{\varepsilon}_k + \bar{u}_k^T \bar{R} \bar{u}_k + J_{k+1} \,\Big|\, \hat{\bar{\varepsilon}}_{k|k} \Bigr\}.
\end{equation}
Using Lemma \ref{lemma:expectation} and standard quadratic minimization techniques, one shows via mathematical induction that the value function admits the quadratic form
\begin{equation} \label{eq:value_function}
J_k = \mathbb{E}\Bigl\{ \hat{\bar{\varepsilon}}_{k|k}^T \Sigma_k \hat{\bar{\varepsilon}}_{k|k} \Bigr\} + \Phi_k,
\end{equation}
with the recursion in \eqref{eq:riccati_recursion}--\eqref{eq:riccati_terminal} and the optimal control law given by \eqref{eq:consensus_protocol_final} and \eqref{eq:optimal_gain}. The complete derivation, which accounts for the contributions of the estimation error, process and measurement noises, as well as the event-triggered communication errors, is omitted here due to space constraints. Interested readers are referred to \cite{20,21} for similar derivations in the deterministic and stochastic settings.
\end{proof}
\begin{remark}
Theorem \ref{thm:main} establishes that, under the proposed event-triggered communication mechanism, an optimal distributed consensus protocol can be designed that minimizes a quadratic cost function while accounting for communication constraints and stochastic uncertainties. This result provides a systematic framework for balancing control performance and communication efficiency in networks of 2D mobile robots.
\end{remark}

\begin{remark}
The result of Theorem~\ref{thm:main} reveals that the design of the LQG consensus controller is decoupled from the design of the event-triggered scheduler. Nevertheless, the quality of the information transmitted—specifically, the magnitude of the triggering error—can adversely affect the overall consensus performance. Thus, while controller synthesis can proceed independently, careful tuning of the event-triggering parameters is essential to ensure robust convergence and satisfactory control performance.
\end{remark}
\section{Stability Analysis}

To assess the stability of the discrete-time system \eqref{eq:global_dynamics} under the optimal consensus protocol, we first introduce a useful lemma from the literature.

\begin{lemma}[\cite{22}] \label{lemma:stability}
Assume there exists a stochastic Lyapunov function \(V_k(\zeta_k)\) and positive constants \(\underline{\kappa}\), \(\bar{\kappa}\), \(\mu>0\), and a scalar \(0<\rho\leq1\) such that
\begin{equation} \label{eq:lyapunov_bounds}
\underline{\kappa}\|\zeta_k\|^2 \leq V_k(\zeta_k) \leq \bar{\kappa}\|\zeta_k\|^2,
\end{equation}
and
\begin{equation} \label{eq:lyapunov_decrease}
\mathbb{E}\{V_k(\zeta_k) \mid \zeta_{k-1}\} - V_{k-1}(\zeta_{k-1}) \leq \mu - \rho\, V_{k-1}(\zeta_{k-1}).
\end{equation}
Then, the stochastic process \(\{\zeta_k\}\) is exponentially bounded in the mean square sense, i.e.,
\begin{equation} \label{eq:mean_square_bound}
\mathbb{E}\{\|\zeta_k\|^2\} \leq \frac{\bar{\kappa}}{\underline{\kappa}} \mathbb{E}\{\|\zeta_0\|^2\}(1-\rho)^k + \frac{\mu}{\underline{\kappa}} \sum_{i=1}^{k-1}(1-\rho)^i,
\end{equation}
and the process is bounded almost surely.
\end{lemma}

We now establish the following result regarding the mean square stability of the consensus error.

\begin{theorem} \label{thm:stability}
Consider the stochastic discrete-time system \eqref{eq:global_dynamics} subject to the event-triggered mechanisms \eqref{eq:etcm1} and \eqref{eq:etcm2}, and the distributed consensus protocol \eqref{eq:consensus_protocol}. Suppose that the estimation error covariance satisfies
\begin{equation} \label{eq:covariance_bound}
P_{k+1|k+1} \leq P_{k+1|k} \leq \bar{p},
\end{equation}
and there exists a positive constant \(\varepsilon^i\) such that
\[
\mathbb{E}\{\|\bar{\varepsilon}_{0|0}\|^2\} \leq \varepsilon^i,
\]
where \(\bar{\varepsilon}_{k|k} = \bar{x}_{k|k} - \mathbf{1}_N \otimes x_0\) denotes the global consensus error. Then, for every robot \(i\) in the network, the consensus error is exponentially bounded in mean square.
\end{theorem}

\begin{proof}
Define the candidate Lyapunov function as
\begin{equation} \label{eq:lyapunov_function}
V_k = \hat{\bar{\varepsilon}}_{k|k}^T \Sigma_k \hat{\bar{\varepsilon}}_{k|k},
\end{equation}
where \(\Sigma_k\) is the positive definite matrix obtained from the Riccati recursion in \eqref{eq:riccati_recursion}. Since \(\Sigma_k\) is positive definite and \(\hat{\bar{\varepsilon}}_{k|k} \neq 0\) when the consensus error is nonzero, it follows that \(V_k\) is a valid Lyapunov function.

We now examine the conditional difference:
\begin{equation} \label{eq:lyapunov_diff}
\begin{aligned}
&\mathbb{E}\{V_{k+1}\mid\hat{\bar{\varepsilon}}_{k|k},\dots,\hat{\bar{\varepsilon}}_{0|0}\} - V_k \\
&\quad = \mathbb{E}\{\hat{\bar{\varepsilon}}_{k+1|k+1}^T \Sigma_{k+1} \hat{\bar{\varepsilon}}_{k+1|k+1} \mid\hat{\bar{\varepsilon}}_{k|k},\dots,\hat{\bar{\varepsilon}}_{0|0}\} - \hat{\bar{\varepsilon}}_{k|k}^T \Sigma_k \hat{\bar{\varepsilon}}_{k|k}.
\end{aligned}
\end{equation}
Using the dynamics in \eqref{eq:global_dynamics_recast} and following standard manipulations (which account for the contributions from the control, communication, and noise terms), we obtain
\begin{equation} \label{eq:lyapunov_decrement}
\mathbb{E}\{V_{k+1}\mid\cdot\} - V_k \leq \mu_k - \rho\, V_k,
\end{equation}
where the term \(\mu_k\) aggregates the trace contributions from the process noise \(\bar{w}_k\), measurement noise \(\bar{v}_{k+1}\), and the event-triggering error \(\bar{e}_k\):
\begin{equation} \label{eq:mu_k}
\scalebox{0.85}{$
\begin{aligned}
\mu_k &= \mathrm{Tr}\Bigl\{(I_N \otimes A)^T(I_N \otimes C)^T\bar{K}_{k+1}^T\Sigma_{k+1}\bar{K}_{k+1}(I_N \otimes C)(I_N \otimes A)\bar{P}_{k|k}\Bigr\} \\
&\quad + \mathrm{Tr}\Bigl\{\bar{K}_{k+1}^T\Sigma_{k+1}\bar{K}_{k+1}\bar{V}\Bigr\} \\
&\quad + \mathrm{Tr}\Bigl\{(I_N \otimes C)^T\bar{K}_{k+1}^T\Sigma_{k+1}\bar{K}_{k+1}(I_N \otimes C)\bar{W}\Bigr\} \\
&\quad + \mathrm{Tr}\Bigl\{(\mathcal{L}\otimes B L_k)^T\Sigma_{k+1}(\mathcal{L}\otimes B L_k)\,\mathbb{E}\{\bar{e}_k^T\bar{e}_k\}\Bigr\}.
\end{aligned}
$}
\end{equation}
By defining a suitable scalar \(\rho>0\) that satisfies
\begin{equation} \label{eq:rho_def}
\begin{aligned}
\rho\, V_k \leq \hat{\bar{\varepsilon}}_{k|k}^T\mathbb{E}\Bigl\{\Bigl[(I_N\otimes A)-(\mathcal{L}\otimes B)L_k\Bigr]^T \\ \times \Sigma_{k+1}\Bigl[(I_N\otimes A)-(\mathcal{L}\otimes B)L_k\Bigr]\Bigr\}\hat{\bar{\varepsilon}}_{k|k},
\end{aligned}
\end{equation}
we can rewrite \eqref{eq:lyapunov_decrement} as
\begin{equation} \label{eq:lyapunov_decrement_final}
\mathbb{E}\{V_{k+1}\mid\cdot\} - V_k \leq \mu_k - \rho\, V_k.
\end{equation}
Since the estimation error covariance is bounded as in \eqref{eq:covariance_bound} and assuming the initial error satisfies \(\mathbb{E}\{\|\bar{\varepsilon}_{0|0}\|^2\} \leq \varepsilon^i\), we can apply Lemma~\ref{lemma:stability} to conclude that
\begin{equation} \label{eq:final_bound}
\mathbb{E}\{\|\bar{\varepsilon}_{k|k}\|^2\} \leq \frac{\bar{\kappa}}{\underline{\kappa}}\mathbb{E}\{\|\bar{\varepsilon}_{0|0}\|^2\}(1-\rho)^{k+1} + \frac{\mu}{\underline{\kappa}} \sum_{m=1}^{k-1}(1-\rho)^m,
\end{equation}
which proves that the consensus error is exponentially bounded in mean square. The almost sure boundedness then follows by standard stochastic stability arguments.

\end{proof}

\begin{remark}
Theorem~\ref{thm:stability} shows that the upper bound on \(\mu_k\) is directly influenced by the process and measurement noise covariances, as well as the trace of the event-triggering error. These factors, in turn, affect the overall consensus error \(\mathbb{E}\{\|\bar{\varepsilon}_{k|k}\|^2\}\). By appropriately tuning the event-triggering condition, one can effectively constrain the upper bound on the consensus error while simultaneously reducing the average transmission rate.
\end{remark}
\begin{algorithm}
\caption{Distributed LQG Rendezvous for 2D Mobile Robots via Event-Triggered Communication}
\label{alg:distributedLQG}
\begin{algorithmic}[1]
\small
\Require 
\begin{itemize}
    \item For each robot \(i \in \{1,2,\ldots,N\}\): initial position \(\boldsymbol{X}_i(0) \in \mathbb{R}^{n^i}\) and sensor reading \(\boldsymbol{Y}_i(0) \in \mathbb{R}^{n^i}\);
    \item Initial state estimates \(\hat{\boldsymbol{X}}^i_{0|0}\) and \(\hat{\boldsymbol{Y}}^i_{0|0}\) with corresponding error covariances \(P^{X_i}_{0|0}\) and \(P^{Y_i}_{0|0}\);
    \item A small tolerance \(\epsilon \ll 0.05\);
    \item System matrices \(A\), \(B\), \(C\); noise covariances \(Q^{X_i}, R^{X_i}, Q^{Y_i}, R^{Y_i}\); and communication weights \(a_{ij}(k)\).
\end{itemize}
\Ensure Updated position estimates \(\boldsymbol{X}_i(k)\), \(\boldsymbol{Y}_i(k)\); speed commands \(v_i^X(k)\), \(v_i^Y(k)\); and performance index \(J^i\).
\While{\(\|\boldsymbol{X}_i(k) - \boldsymbol{X}_j(k)\| > \epsilon\) \textbf{or} \(\|\boldsymbol{Y}_i(k) - \boldsymbol{Y}_j(k)\| > \epsilon\)}
    \State \textbf{Measurement Update:} Obtain sensor measurements (e.g., odometry, IMU) and update state estimates using the Kalman filter:
    \begin{align*}
      \hat{\boldsymbol{X}}^i_{k|k} &= A\,\hat{\boldsymbol{X}}^i_{k|k-1} + B\,u^X_{i}(k-1) + K^X_i(k)\Bigl[z^X_i(k) - C\,\hat{\boldsymbol{X}}^i_{k|k-1}\Bigr],\\[1mm]
      P^{X_i}_{k|k} &= \bigl(I - K^X_i(k)C\bigr) P^{X_i}_{k|k-1},
    \end{align*}
    and similarly update \(\hat{\boldsymbol{Y}}^i_{k|k}\) and \(P^{Y_i}_{k|k}\).
    
    \State \textbf{Event-Triggered Transmission:} 
    \begin{itemize}
      \item Check the event-triggering condition (cf. \eqref{eq:etcm1} and \eqref{eq:etcm2}). 
      \item If the condition is satisfied, robot \(i\) broadcasts its current state estimates \(\hat{\boldsymbol{X}}^i_{k|k}\) and \(\hat{\boldsymbol{Y}}^i_{k|k}\) to its neighbors.
    \end{itemize}
    
    \State \textbf{Consensus Gain Computation:} 
    \begin{itemize}
      \item Solve the discrete-time Riccati equations to obtain \(\Sigma_k^X\) and \(\Sigma_k^Y\):
      \begin{align*}
        \Sigma_k^X &= (A - B L_k^X)^T \Sigma_{k+1}^X (A - B L_k^X)  + L_k^{X^T} R^X L_k^X + Q^X,\\[1mm]
        \Sigma_k^Y &= (A - B L_k^Y)^T \Sigma_{k+1}^Y (A - B L_k^Y)  + L_k^{Y^T} R^Y L_k^Y + Q^Y,
      \end{align*}
      with terminal conditions \(\Sigma_M^X = Q_M^X\) and \(\Sigma_M^Y = Q_M^Y\).
      \item Compute the optimal consensus control gains:
      \begin{align*}
        L_k^X &= \Bigl[R^X + B^T\Sigma_{k+1}^X B\Bigr]^{-1} B^T\Sigma_{k+1}^X A,\\[1mm]
        L_k^Y &= \Bigl[R^Y + B^T\Sigma_{k+1}^Y B\Bigr]^{-1} B^T\Sigma_{k+1}^Y A.
      \end{align*}
    \end{itemize}
    
    \State \textbf{Consensus Update:} For each robot \(i\), update the positions using the distributed consensus protocol:
    \begin{align*}
      \boldsymbol{X}_i(k+1) &= \boldsymbol{X}_i(k) - L_k^X \sum_{j \in \mathcal{N}_i} a_{ij}(k) \Bigl[\hat{\boldsymbol{X}}^i_{k|k} - \hat{\boldsymbol{X}}^j_{k|k}\Bigr],\\[1mm]
      \boldsymbol{Y}_i(k+1) &= \boldsymbol{Y}_i(k) - L_k^Y \sum_{j \in \mathcal{N}_i} a_{ij}(k) \Bigl[\hat{\boldsymbol{Y}}^i_{k|k} - \hat{\boldsymbol{Y}}^j_{k|k}\Bigr].
    \end{align*}
    
    \State \textbf{Compute Performance Index:} Evaluate the performance index for robot \(i\):
    \[
    J^i = \mathbb{E}\left[\hat{\varepsilon}_M^{i^T} Q_M^i \hat{\varepsilon}_M^i + \sum_{k=0}^{M-1} \left(\hat{\varepsilon}_k^{i^T} Q^i \hat{\varepsilon}_k^i + u_k^{i^T} R^i u_k^i\right)\right],
    \]
    where \(\hat{\varepsilon}_k^i = \hat{\boldsymbol{X}}^i_{k|k} - x_0\) (or similarly for the \(Y\) component) represents the consensus error.
    
    \State Update \(k \gets k+1\).
\EndWhile
\end{algorithmic}
\end{algorithm}

\section{Experimental validation}
\begin{figure}[h] 
  \begin{center}
\includegraphics[width=0.45 \textwidth]{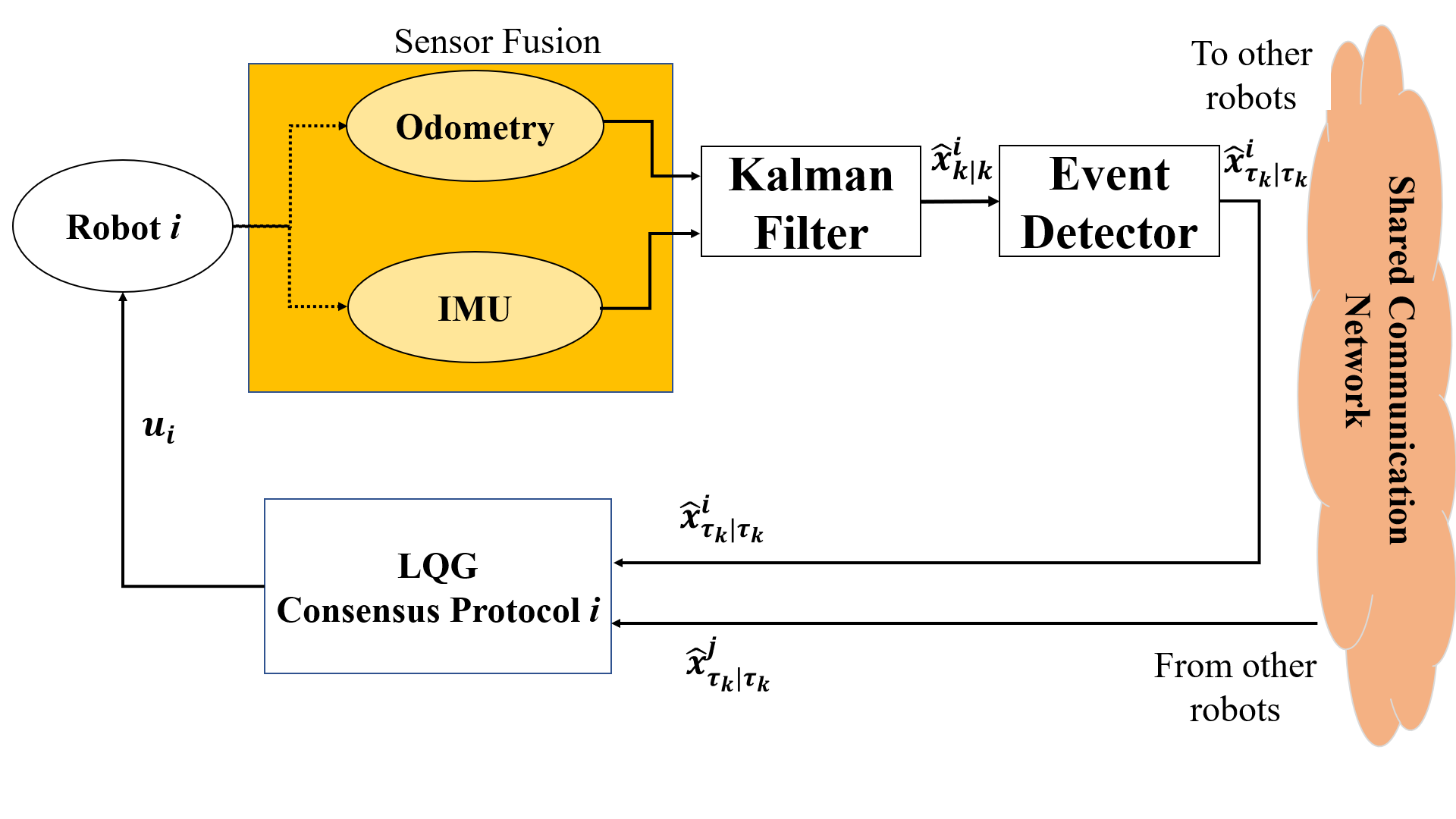}
  \end{center}
  \caption{Diagram of proposed distributed LQG consensus optimal control with event-triggered scheduler } 
  \label{Fig3}
\end{figure}

\begin{figure}[h!] 
  \begin{center}
\includegraphics[width=0.4 \textwidth]{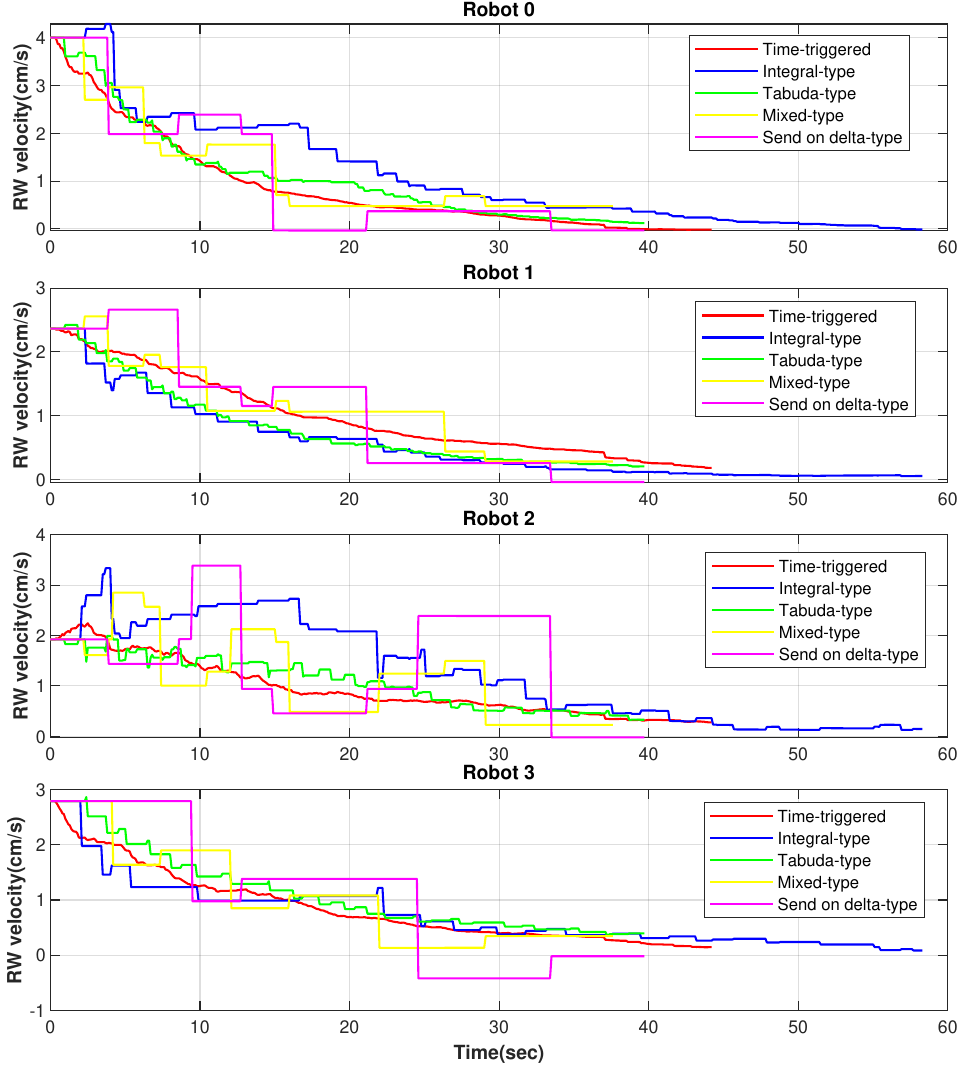}
  \end{center}
  \caption{Experimental testing results of the proposed LQG consensus control: right wheel velocities of four  mobile robots considering different  triggering conditions } 
  \label{Fig3}
\end{figure} 

\begin{figure}[h!] 
  \begin{center}
\includegraphics[width=0.4 \textwidth]{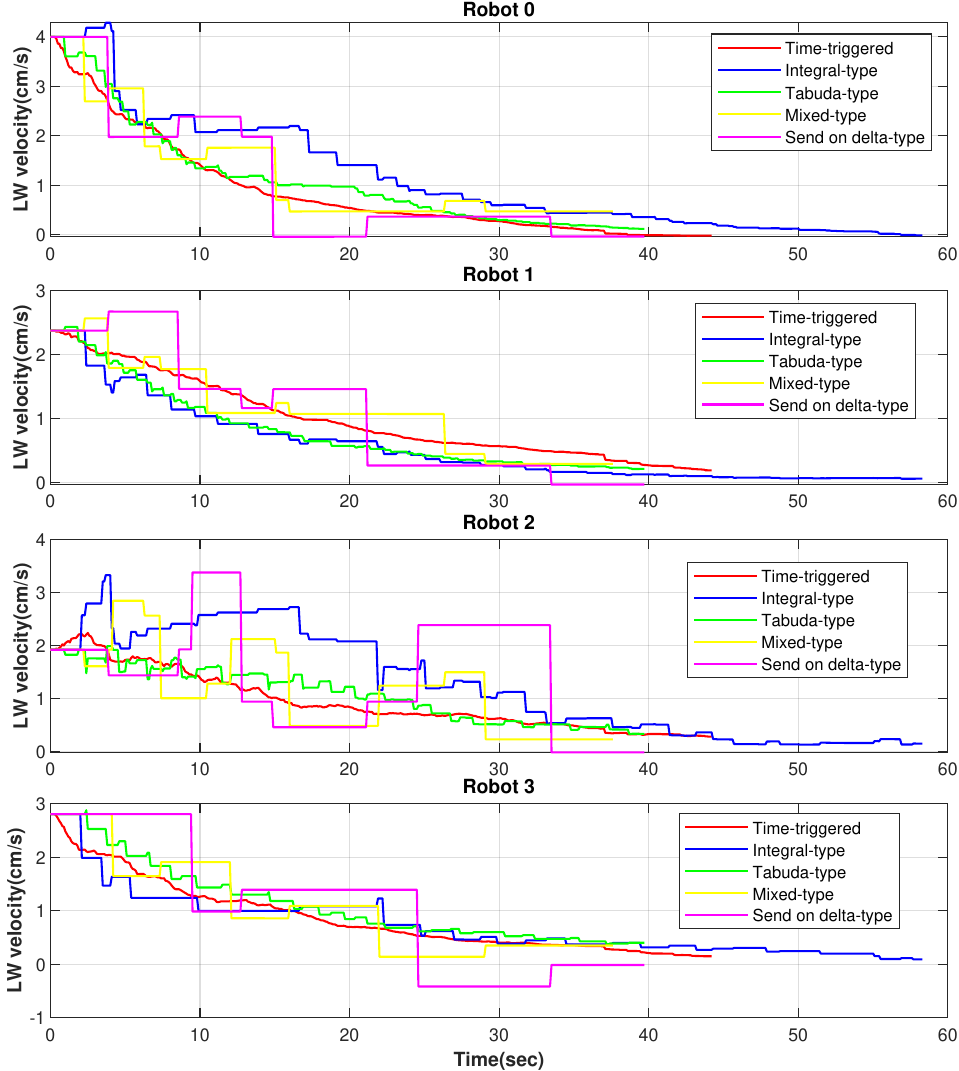}
  \end{center}
  \caption{Experimental testing results of the proposed LQG consensus control: left wheel velocities of four  mobile robots considering different  triggering conditions } 
  \label{Fig3}
\end{figure} 
\begin{figure}[h] 
  \begin{center}
\includegraphics[width=0.4 \textwidth]{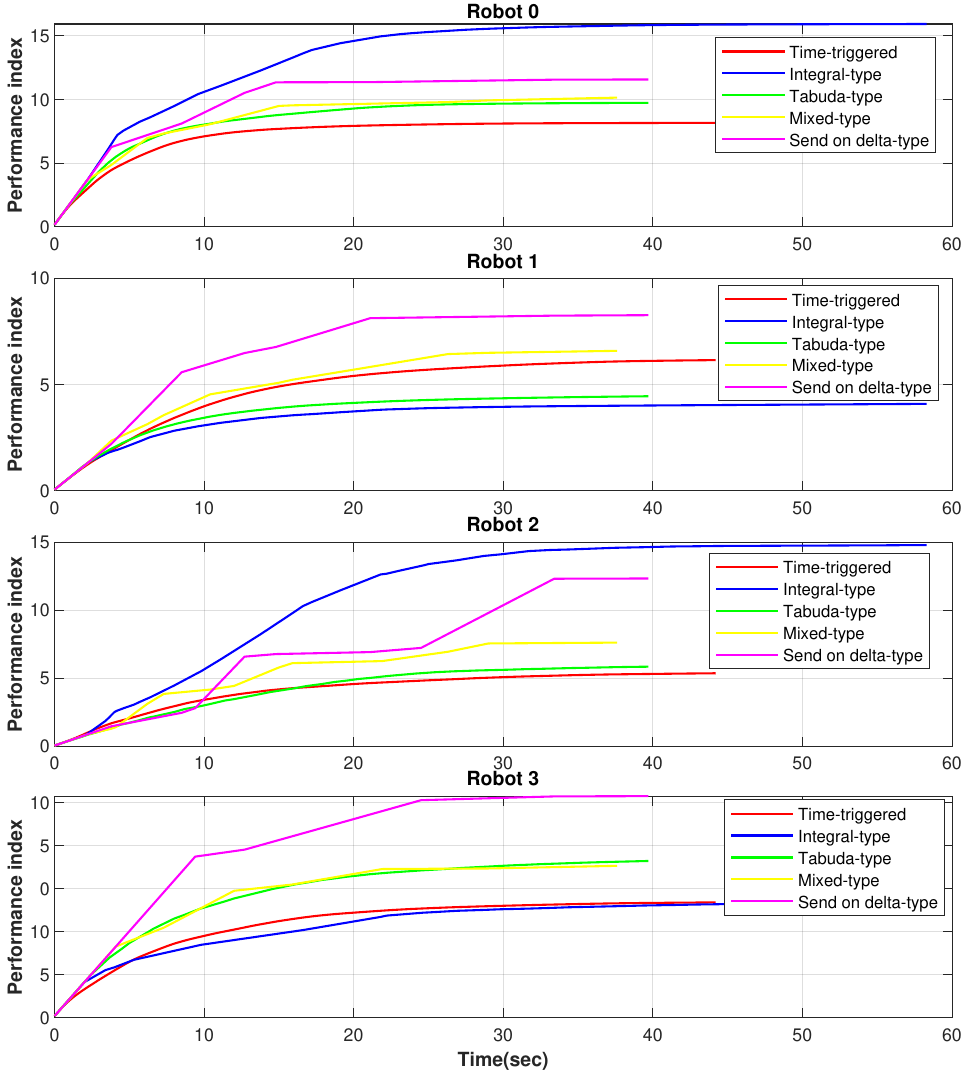}
  \end{center}
  \caption{Experimental testing results of the proposed LQG consensus control: performance index of four mobile robots considering different triggering conditions} 
  \label{Fig3}
\end{figure} 

In this section, we experimentally evaluate the performance of the proposed LQG rendezvous control with an event-triggered mechanism using a team of Robotrium mobile robots. The Robotrium robots operate in a two-dimensional plane, where the coordinates \((X, Y)\) represent their Cartesian positions. This workspace frame is used to control the robots. The heading \(\theta\) denotes the angle between the positive \(Y\)-axis and the direction in which the front of the Robotrium is facing. The angle \(\theta\) is measured counterclockwise from the positive \(X\)-axis, with values ranging from \(-180^\circ\) to \(180^\circ\). Assuming decoupled motion along the \(X\)- and \(Y\)-axes, the system dynamics of the Robotrium robots along these axes can be expressed as follows:

\begin{equation}
\begin{aligned}
\dot{X}_i(t) &= u_X^i(t) = v_i(t)\cos(\theta_i) \\
\dot{Y}_i(t) &= u_Y^i(t) = v_i(t)\sin(\theta_i), \quad i = 1, \dots, 4
\end{aligned}
\end{equation}

where \(X_i(t), Y_i(t) \in \mathbb{R}\) represent the position of robot \(i\), and \(u_X^i(t), u_Y^i(t) \in \mathbb{R}\) are the control inputs along the \(X\)- and \(Y\)-axes, respectively. The term \(v_i(t)\) is the linear velocity of the robot.

Figure 1 depicts the proposed LQG rendezvous control strategy for mobile robots. In this setup, each robot samples and broadcasts its noisy position information to neighboring robots to design the optimal rendezvous protocol. An Extended Kalman Filter (EKF) is employed to estimate the positions of the robots. Each Robotrium robot decides whether to transmit its most recent position estimate to its neighbors.  Each robot employs the EKF to estimate its position based on odometry and IMU sensor data.

The initial positions of the robots are as follows:
\[
x_{\text{Robot 0}}(0) = 0.2, y_{\text{Robot 0}}(0) = -0.1,
\]
\[
x_{\text{Robot 1}}(0) = -0.2, y_{\text{Robot 1}}(0) = -0.1,
\]
\[
x_{\text{Robot 2}}(0) = -0.2, y_{\text{Robot 2}}(0) = 0.1,
\]
\[
x_{\text{Robot 3}}(0) = 0.2, y_{\text{Robot 3}}(0) = 0.065,
\]
with zero initial velocities. Additionally, the initial position estimates of the robots are assumed to be identical to their actual initial positions.

The goal of the experiment is to design and implement the LQG rendezvous protocol under various triggering conditions to ensure that the Robotrium robots achieve rendezvous while optimizing cost-energy performance. We compare the positions, velocities, and performance indices using the LQG rendezvous protocol under five types of triggering conditions: time-triggered, integral-type, Tabuada-type, mixed-type, and send-on-delta-type. Figures 2, 3, and 4 show the rendezvous results for the positions, right wheel velocities, left wheel velocities, and performance indices of the Robotrium robots under these different triggering conditions. Tables I and II present the average transmission rates for each robot along both the \(X\)- and \(Y\)-axes under each triggering condition.

The results indicate that the time-triggered mechanism provides the minimum cost performance when compared to the other triggering conditions, but it transmits unnecessary information to the controller. On the other hand, the send-on-delta mechanism yields the maximum control cost performance but transmits the least amount of information to the controller. Based on these findings, we conclude that by carefully tuning the event-based scheduler, we can achieve satisfactory rendezvous performance while reducing the average transmission rate.

\begin{table}[h]
	\renewcommand{\arraystretch}{1.}
	\captionsetup{justification=centering}
	\caption{Average transmission rate for different event-triggered conditions considering the motion of robot along x-axis }
	\centering
	\label{table_1}
	\resizebox{\columnwidth}{!}{
		\begin{tabular}{l l l l l}
			\hline\hline \\[-3mm]
			\multicolumn{1}{c}{Type of e-puck robot} & \multicolumn{1}{c}{Robot 0}  & \multicolumn{1}{c}{Robot 1} & \multicolumn{1}{c}{Robot 2} & \multicolumn{1}{c}{Robot 3} \\[1.6ex] \hline
			 Time-triggered  & 0.57  & 0.55 & 0.54 & 0.57 \\ \hline
			 Integral-type  & 0.06  & 0.04 & 0.02 & 0.03 \\ \hline
			 Tabuda-type  & 0.11  & 0.07 & 0.04 & 0.05 \\ \hline
			 Mixed-type  & 0.01  & 0.01 & 0.01 & 0.01 \\ \hline
			 Send on delta-type  & 0.007  & 0.007  & 0.007  & 0.007
			 \\
			\hline\hline
		\end{tabular}
	}
\end{table}

\begin{table}[ht!]
	\renewcommand{\arraystretch}{1.}
	\captionsetup{justification=centering}
	\caption{Average transmission rate for different event-triggered conditions considering the motion of robot along y-axis }
	\centering
	\label{table_1}
	\resizebox{\columnwidth}{!}{
		\begin{tabular}{l l l l l}
			\hline\hline \\[-3mm]
			\multicolumn{1}{c}{ Type of e-puck robot } & \multicolumn{1}{c}{Robot 0}  & \multicolumn{1}{c}{Robot 1} & \multicolumn{1}{c}{Robot 2} & \multicolumn{1}{c}{Robot 3} \\[1.6ex] \hline
			 Time-triggered  & 0.71  & 0.63 & 0.7 & 0.63 \\ \hline
			 Integral-type  & 0.07  & 0.07 & 0.07 & 0.008 \\ \hline
			 Tabuda-type  & 0.13  & 0.12 & 0.13 & 0.01 \\ \hline
			 Mixed-type  & 0.04  & 0.04 & 0.05 & 0.01 \\ \hline
			 Send on delta-type  & 0.05  & 0.05  & 0.07  & 0.03
			 \\
			\hline\hline
		\end{tabular}
	}
\end{table}

\section{Conclusion}
In this paper, we presented a novel LQG rendezvous control framework for multi-robot systems operating under localization uncertainty. By integrating an event-triggered communication mechanism with the optimal control design, our approach effectively reduces communication load while ensuring robust and efficient rendezvous performance. The design decouples the control law from the event-triggered scheduler, allowing independent tuning of both the controller and the communication parameters. Experimental validation on a team of Robotrium mobile robots demonstrated that the proposed protocol achieves satisfactory performance under various triggering conditions, with a clear trade-off between control cost and transmission rate. Future work will focus on extending the framework to more complex robotic dynamics and heterogeneous networks, as well as exploring adaptive strategies for dynamic event-triggering in rapidly changing environments.

\end{document}